**Understanding Energy Flow and Inefficiency of a Thermomagnetic Generator by Transient Multi-Physics Modelling**

Ali Izadi[1,2, a)], Bruno Neumann[1,2], Sebastian Fähler[1]

1) Helmholtz-Zentrum Dresden-Rossendorf, Bautzner Landstraße 400, 01328 Dresden, Germany
2) Faculty of Mechanical Science and Engineering, TUD Dresden University of Technology, 01062 Dresden, Germany

# Abstract

Waste heat recovery improves energy efficiency and reduces greenhouse gas emissions; however, much industrial and environmental heat is wasted at low temperature. Thermomagnetic recovery of waste heat has a high potential for sustainable production of electric energy, especially for low-grade waste heat where conventional technology is inefficient or infeasible. Of particular interest are thermomagnetic generators (TMG) as they require almost no mechanically moving parts, which is beneficial for high reliability. However, all existing prototypes have two remaining challenges: low efficiency and low cycle frequency. In this work, we develop a digital twin of a recent TMG with genus 3 by using multi-physics simulations. We identify shortcomings of previous simulation approaches, and describe why simulations in three dimensions are necessary, which consider coupling between magnetic, thermal, fluid flow, and electrical physics domains. We validate our model, which only uses known geometry and material parameters, by experimental data of the TMG with highest power density today, and attain 96% accuracy in open-circuit voltage and 95% accuracy in power output. This high accuracy allows us to identify the origin of both challenges for TMGs, which are not accessible by experiments. First, we uncover inefficiencies by analyzing the energy flow within a Sankey diagram. Second, we trace the transient heat flow through the generator, which identifies the factors limiting frequency. This paves the way for more efficient and faster TMGs, and their development will be accelerated by our validated digital twin.

a) a.izadi@HZDR.de

# 1. Introduction

Recovering energy from low-grade waste heat is increasingly important for improving industrial energy efficiency and reducing emissions [1] [2]. Aligned with sustainable development goals, reducing energy losses today helps preserve resources and environmental capacity for future generations[3]. Studies estimate more that about half of global energy consumption is ultimately lost as waste heat [1] [2]. A large portion of this waste heat arises at low temperatures, for example around 60–65% of industrial waste heat is below 100 °C [4] [5]. Various technical and economic barriers still hamper the exploitation of low-grade heat, including thermodynamic limits, dispersed sources, and cost of recovery  [6]. This motivates the development of viable methods to harvest electricity from low-grade waste heat, which yields substantial energy and cost savings [1]. Several low-grade heat recovery technologies are under exploration, for instance, heat engine cycles like Organic Rankine Cycles (ORCs) can produce power from moderate temperature differentials [7]. Also, solid-state techniques such as Thermoelectric generators (TEGs)[8] , which using the Seebeck effect, are under strong development, but have a low intrinsic energy density at small temperature differences and require harmful materials [9]. These drawbacks have motivated the search for alternative low-grade heat harvesting technologies, which include thermoacoustic[10], thermoelastic [11], pyroelectric [12], and thermomagnetic systems [13].

Of particular interest are thermomagnetic systems, which is a promising technology for harvesting waste heat below 100 °C[15] since this technology shows a high thermodynamic efficiency. Calculations, based on material properties indicate that efficiency can be up to 55% of the Carnot limit [14]. Since the Carnot limit of any system operating e. g. between 300 K and 330 K, is around 10%, TMGs should be able to convert up to 5.5% of this low-grade waste heat to electricity.

In this work we focus on Thermomagnetic Generators (TMG), as this particular thermomagnetic system requires almost no mechanically moving parts, which promises low maintenance and long lifetime. Within a TMG, the TMM is cyclically heated and cooled through its Curie temperature. At this $T_C$, the TMM exhibits a sharp change of magnetization in a narrow temperature range. This allows to switch the magnetic flux, created by a permanent magnet, and guided by soft magnetic yokes. When alternating the temperature of the TMM between waste heat $T_{hot}$ and ambient $T_{cold}$, an additional induction coil converts this flux change to electricity according to Faraday's law of induction. It's worth mentioning that all existing demonstrators, described in more detail later, reach a system efficiency, which is orders of magnitude lower compared to the value given above. To make this difference clear, we distinguish between *system* efficiency, which is the ratio between electric output energy and waste heat input energy measured for a demonstrator system, from the *materials* efficiency, estimated from the material properties. The second important performance criterium for a TMG is the output power, which is equal to the energy converted per cycle times frequency. While the energy per cycle depends on efficiency and system size, a high frequency has the advantage, that it also increases the power density. Indeed, the cycle frequencies of today's demonstrator systems is below 1 Hz, thus there should be plenty of room to become faster.

Here we aim to understand the origins of the gap between system and material efficiency, and of the low cycle frequency of TMGs. For this we develop a digital twin of an experimentally fully characterized TMG[13]. We analyse the shortcomings of previous simulations and explain how these can be improved by transient 3-dimensional multiphysics simulation, which consider coupling of magnetic and thermal subsystems. Then we validate our simulations by the experimentally measured open circuit voltage and output power. Finally, we use our digital twin to identify the origins of two major shortcomings of today's TMGs which are not accessible by experiments: The low thermodynamic efficiency and the low cycle frequency, which both limit the electrical output power. First, however, a short review on TMGs and their simulation is required, which summarizes the solved and open questions in more details.

# 2. Performance and simulation of thermomagnetic generators prototypes

In the following we sketch the developments and simulations of TMG in historical sequence. This allows us to understand the improvements of electrical output power and frequency with time, depicted in **Figure 1**. We do not compare efficiency, since in papers often only unrealistic high estimations from materials efficiency are given, but not a measured device efficiency. We consider this as not appropriate for comparing devices. Instead, we extracted the power per active material volume and report this power density in **Figure 1.c**. Scaling power to the amount of active material, is necessary since device size varies strongly, and power density describes which design makes best use of the expensive active material. We list only TMGs, as they do not require moving parts, which is of benefit for low maintenance. Other thermomagnetic systems like thermomagnetic motors [16], microsystems [17], and oscillators [18], which involve mechanical movements, are not included.

The first concepts of TMGs were proposed in the late 19$^{th}$ century in patents of Edison [19] and Tesla [20], but there is no report on an experimental realization. Calculations dating back to Brillouin and Iskenderian in 1948 showed that thermomagnetic system could achieve high efficiency [21]. Though their analytical analysis describes materials and not system efficiency, their work was a breakthrough, identifying the high potential of TMGs for low-grade waste heat recovery. Then for many years, the experimental development was hindered by the lack of suitable materials: Neither TMM with Curie temperatures near room temperature and steep magnetization change where available, nor sufficiently strong permanent magnets.

During the strong progress in material science within the 21$^{st}$ century several new functional alloys were discovered, including magnetic shape memory alloys [22]. The multiferroic properties of these Heusler alloys enabled Srivastava et al. [23] in 2011 to reconsider TMGs and they developed a simple setup to convert heat directly into electricity. For their proof of principle, they used a bulky sample with 7.5 mm diameter, which makes heat exchange slow and accordingly only a frequency of 0.1 Hz was reached. Wrapping a coil around the TMM, allowed to measure an open voltage peak of 0.6 mV, though this setup did not have any closed magnetic circuit.

The next progress was also obtained from a material science group, which worked on magnetocaloric materials, developed for the reverse energy conversion process. In 2014 Christiaanse and Brück [24] presented a novel TMG prototype, which alternatingly heats and cools two TMM, which allows to switch the magnetic flux between two closed loops, avoiding stray fields. They wrapped the induction coils around the TMM consisting of $(Mn,Fe)_2(P,As)$, which required to make holes into this material for fluid flow. Since machining of most TMM is difficult, device assembly appears complex, and probably therefore a large difference in theoretical and actual performance occurred. This device had a frequency of 0.17 Hz, and generated a maximum power of 0.045 mW. They optimized the magnetic circuit using a magnetostatic Radia Mathematica simulation with measured magnetic properties and predicted the mean field and field change through simulations. However, the realized prototype achieved a field change several times less than predicted value, indicating substantial practical losses because the active material was affected by laser drilling.

A significant breakthrough came in 2019 when Waske et al. [13] presented a pretzel-like magnetic field topology, which in addition to avoiding stray fields, allowed for a reversal of flux direction. They attributed the strong improvement of TMG's performance to this, and achieved a peak output power of 0.8 mW at optimum frequency of 0.8 Hz. They reached about 50% of the simulated flux change, though only static 2D magnetic simulations were used. Indeed, this work contained a further design improvement. They realized, that there is no need to wrap the coil around the TMM, but it's sufficient to wrap it around the yoke, guiding the same flux. This gives direct access to the TMM, and accordingly thin La–Fe–Co–Si plates were used, which allow for a fast heat exchange. This work serves as the experimental reference for the present study.

In 2021, Dzekan et al. [25] used the same TMG setup and examined the influence of the TMM used. When comparing La–Fe–Co–Si with Gadolinium, often used as "standard" material, they found a substantial decrease of power output from 0.8 to 0.12 mW, but a slight increase of optimum frequency from 0.8 to 1.1 Hz. This difference in behavior arises from the sharper temperature dependence of magnetization in La–Fe–Co–Si on one side, and the better heat transfer properties of gadolinium on the other, and illustrates the importance of the TMM material used [14]. This aspect must be neglected in the present work focusing on device engineering, since no other device was tested with different materials.

Jiang et al. [26] used the same magnetic field topology and Gd plates, and expanded TMGs for the first time by a regenerator, which is a well-known concept from magnetocaloric to enhance efficiency and temperature span [27]. To optimize their design, they used COMSOL Multiphysics 5.5 and split the simulation into two parts. One part was a 2D laminar flow and heat transfer model with a moving mesh, and the other part was a 3D magnetic field and electrical circuit model, rather than a fully coupled 3D model; however, the simulated results differ from experiments by orders of magnitude. A maximum power of 25 nW was obtained at a frequency of 0.4 Hz.

More recently, in 2023 Liu et al. [28] proposed another magnetic field topology, which allowed for flux reversal. In contrast to [13], this approach does not avoid magnetic stray fields (as evident from **Figure 2.a** [28]). Gd plates were used as TMM, and at optimum conditions of 0.25 Hz a power output of 123.3 µW was delivered. They try to illustrate this by lightning a LED, but indeed the generated power was amplified, as described only in the supplementary. They used a COMSOL finite element Multiphysics model, coupling heat transfer and magnetic field to simulate the induced voltage and compare it to experiments. These simulations were used to optimize device performance by a parametric study.

Finally, in 2024 Bahl et al. [29] developed an upscaled TMG using packed Gd spheres in the classical magnetic field topology [24]. They successfully increased output power to 9 mW at 0.75 Hz, which is the highest value reached by TMGs today. However, this required a quite large system with a lot of TMM. It's worth mentioning that they used COMSOL to optimize their design and maximize the useful flux swing before fabrication. They simulated with a 3D finite-element model, but only addressed magnetostatics using the magnetic scalar potential formulation.

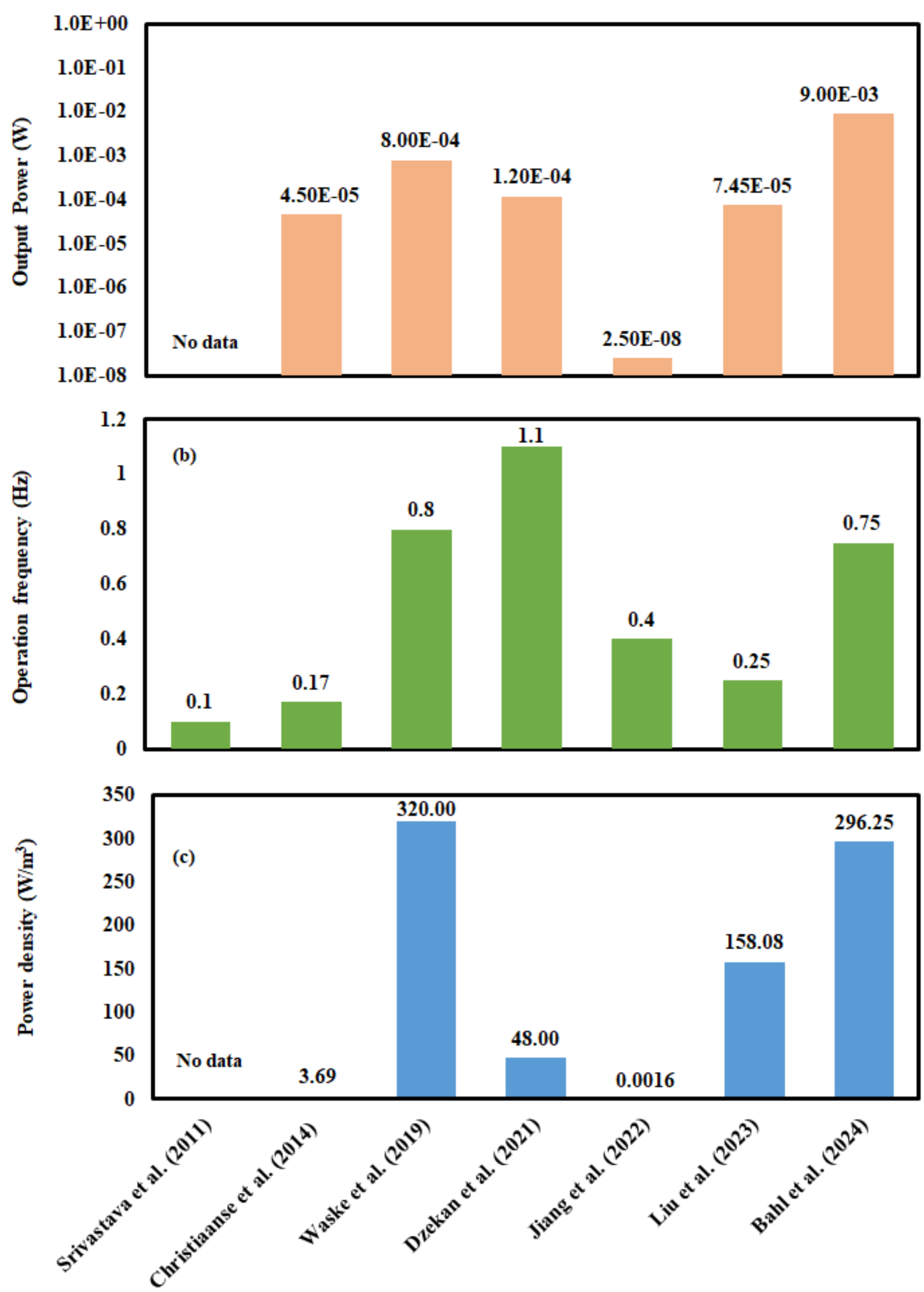

***Figure 1****: Comparison of thermomagnetic generator prototypes sorted by publication time with respect to their experimental a) output power, b) operation frequency and c) power density. Power density of each device is obtained by dividing the power output by the volume of the active thermomagnetic material and is a decisive measure how this device makes use of this expensive material.*

Though all these systems not only differ in design, but also in the TMM material and shape used, it's instructive to compare their performance with respect to power output, optimum operational frequency, and power density (**Figure 1**). All performance indicators only roughly increase with publication time, and there is substantial scatter specially in output power, requiring a logarithmic scale in **Figure 1.a**. We take this as an indication, that there is no agreement for optimum design guidelines yet. Moreover, the few reported values for system efficiency are much lower than material efficiency, which is probably the reason, why this important property of any heat engine is rarely given. For instance, one of the papers that reported this value is [13] which is the experimental reference of the current study. This device has the highest power density, but directly measured device efficiency is just $1.7 \times 10^{-3}$ %.

Though almost all publications use different types of simulations to optimize their device, often there is a large difference to experimental values, which must be understood. Thus, we will first identify requirements for realistic simulations of TMG. Indeed, we do not aim for a parametric device optimization (yet), but for a fundamental understanding of the shortcoming of today's designs. For this, transient simulations of a TMG are necessary, which were not attempted before, probably due the complex coupling of magnetic, fluid and temperature domains.

# 3. TMG simulated

For our simulations we selected the TMG developed by Waske et al. [13], as it reaches the highest power density, and a detailed description of setup and experimental data is available. Accordingly, for details we refer to the original paper and supplementary table S1, which summarizes materials and their sizes. In the following, we just give a brief summary of the function principle.

The function principle is illustrated in **Figure 2**. This TMG contains two sets of TMM, consisting of La–Fe–Co–Si plates with a Curie temperature of 300 K. These sets are heated and cooled alternatingly, to switch the magnetic flux off and on, respectively, as sketched in 2.a) and 2.b). The magnetic flux is created by two Nd-Fe-B permanent magnets, which both have the same magnetization direction and soft magnetic steel yokes guide this flux, as sketched by the yellow arrows. The induction coil is placed between both permanent magnets, where the direction of the magnetic flux is reversed when alternating the temperature between both TMM. This is due to the particular topology of the magnetic circuit, which has a genus of 3. The induced voltage in the coil is given by Faraday's law of induction, where $V_{OC}$ is the open circuit voltage, $N$ represents the number coil turns, Φ is magnetic flux and $t$ is time.

$$V_{OC} = -N \times \frac{d\Phi}{dt} \quad (1)$$

For the simulations we use the experimentally optimized process parameters for maximum electrical output power $P_{el}$, which is obtained by measuring the induced voltage $V$ at a load resistance of $R_l$ = 9.2 Ohms.

$$P_{el} = \frac{V^2}{R_l} \quad (2)$$

Maximum power was obtained for water temperatures of $T_{hot}$= 315 K and $T_{cold}$ = 285 K, frequency $f$= 0.8 Hz at a volumetric flow rate $\dot{Q}$ = 0.4 l/min for each side, which are kept constant during all simulations. The required waste heat $P_{in}$ is measured from the water mass flow rate $\dot{m}$ and the known fluid specific heat capacity $C_p$ of water:

$$P_{in} = \dot{m} \times C_p \times (T_{hot} - T_{cold}) \quad (3)$$

The ratio of both directly gives the system efficiency:

$$\eta_{sys} = {}^{P_{el}}/_{P_{in}} \quad (4)$$

# 4. 2D magnetic physics modelling and its shortcomings

As summarized in section 2, up to now most of the reported simulations were based on 2D planar finite element models (FEM). only the simulations reported in [26] and [29] employed 3D models, but only for one physical domain. Moreover, time dependent simulations of a thermomagnetic generator, which consider coupling between the different domains and physics are missing. In the following we will identify the fundamental shortcomings of previous approach for the present TMG and describe, how they can be overcome. We are aware, that readers from simulation community and TMG engineering have different interests, thus we make intensive use of a supplementary, which contains all details of the COMSOL Multiphysics simulation and results, and just give a compressive summary.

Time dependency captures transient states between completely hot and cold. Accordingly, also data for the TMM at intermediate temperatures is required. As also the magnetic field *B* varies at intermediate stages, complete magnetization data *B(H, T)* is required, which considers the thermomagnetic coupling at material level. As shown and described in supplementary Figure S1, we measured this data by VSM, corrected the field for sample geometry by using the Demag factor, and fitted it by an appropriate arctangent function. This fit is decisive for robust numerical simulations, as it removes all negative slops present in experimental data, which represent a physical and therefore also numerical instability. Thus, all following simulations use this fitted data.

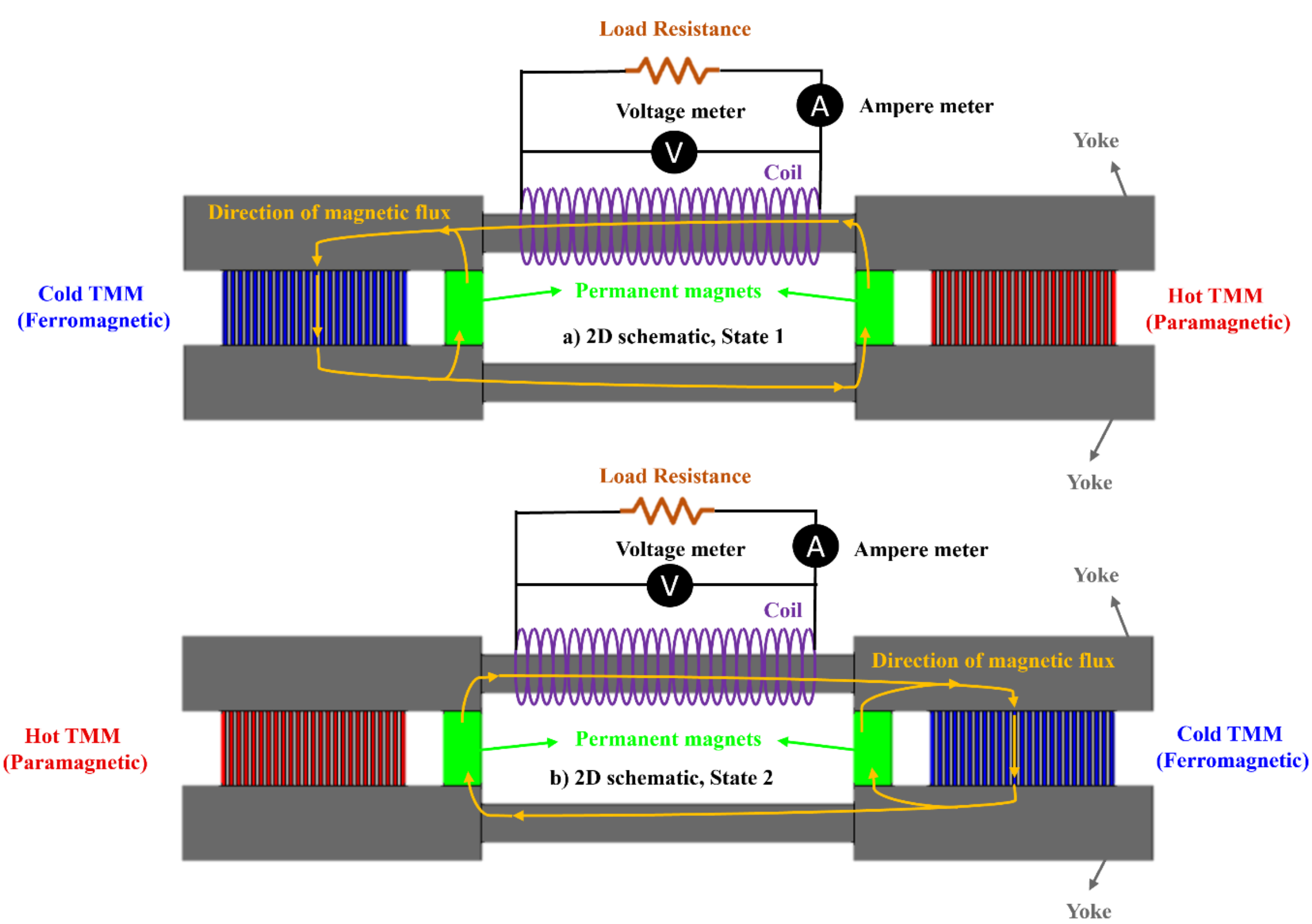

***Figure 2: 2D geometry schematic and working principals of Genus 3 TMG device:*** *The device consists of connecting yokes, two permanent magnets, hot TMMs in paramagnetic state and cold TMMs in ferromagnetic state (Details in supplementary, Table S.1). The TMM plates are heated and cooled by water flow perpendicular to this sketch. A coil is simulated around top yoke, which is connected to a load circuit. There are two states:* ***a)*** *State 1, when the left cold TMMs are in the ferromagnetic state and right hot TMMs are blocking flux in their paramagnetic state and flux goes from right to left and* ***b)*** *state 2, when the left hot TMMs are at paramagnetic state and right cold TMMs are passing the flux in their ferromagnetic state and flux goes from left to right. The device switch at a frequency of 0.8 Hz between states a) and b).*

We use these material data for 2D simulations of our TMG, which is the common approach to optimize parameters quickly, as even a fine mesh can run fast due to the low computational cost. However, in a 2D TMG no fluid flow is possible, as the fluid flows perpendicular to the 2D cut (sketched in **Figure 2**). Accordingly, by principle the heat transfer domain cannot be simulated realistic in 2D. Instead, one has to simplify this, and we use the first order approximation, that the temperature of all TMM varies sinusoidal between hot and cold temperatures, as shown in Figure S2.a for the optimum $f$= 0.8 Hz. Under this 2D configuration, we obtain an oscillating magnetic flux with higher values compared to the 3D simulations, as illustrated in supplementary Figure S2.b. These differences between 2D and 3D simulations occur due to the fact, that a magnetic field relaxes to a minimum energy configuration, which includes avoiding stray fields as much as possible. As shown in Figure S3, in 3D there are additional possibilities to minimize stray fields, which are not accessible in 2D. As stray field energies represent an energy barrier during a TMG cycle [13], even when considering the magnetic domain only, 3D calculations are much more accurate.

Finally, we illustrate the need to considering coupling between the magnetic and fluid domains. If we compare the magnetic flux change in Figures S2.b and S4.a of the supplementary, in non-coupled single physics simulation we observe much more rectangular change of magnetic flux density with time, compare to the experimentally observed sinusoidal curve. This illustrates that the simulated heating and cooling in non-coupled systems is much faster than in experiments. To understand the origin of the low frequency in a TMG therefore 3D multiphysics simulations are required, which can consider coupling between the magnetic and fluid mechanic domain.

To summarize, previous modelling has three shortcomings. First, mostly 2D models were used, which are a priori incomplete for a 3D system and therefore cannot capture geometrically dependent phenomena of the device, such as perpendicular fluid flow and magnetic flux. Second, coupling between the different domains was simplified or even neglected, but this is decisive for TMGs using functional materials, which have coupling already in their name: thermomagnetic. Third, and most importantly, transient calculations are required to follow the flow of heat, which decides on both challenges of TMGs: efficiency and frequency.

# 5. 3D fully coupled transient Multiphysics Simulation

To accurately model the transient TMG's behaviour and capture the impact of 3D effect, we developed a 3D model that couples electromagnetic, thermal, and fluid flow physics as it is shown in **Figure 3**. Simulations were performed using COMSOL Multiphysics (version 6.3) due to its ability to handle coupled field problems.

The physic domains contain: 1) Laminar flow, which is used to simulate the fluid flow along the TMM plates. As in the experiment, water flow is used to switch the plate temperature. In the model, we included a fluid domain in contact with the plates and solved the equations for laminar flow. This allows the model realistic fluid travel time, and heating and cooling rates of the plates. 2) Heat transfer in solids and fluids that solve the transient mixing, convective and conductive heat transfer between the fluid and different parts of the device. In particular heat conduction in the plates, yokes, and magnets, with appropriate thermal boundary conditions to emulate heating/cooling. 3) Magnetic fields (AC/DC) for the calculation of magnetic vector potential in the presence of nonlinear magnetic materials, including the temperature dependent thermomagnetic plates, soft iron yokes, permanent magnet regions, as well as the stray fields outside of the device. 4) Electrical circuit is used to couple the induced voltage within the coil to an external circuit. This interface allows us to connect the coil either to an open circuit, which describes the connection to a voltmeter with a very high internal resistivity, or to a finite load resistance, where the product of current and voltage gives the output power. Essentially, the coil as part of this circuit is modelled as a multi-turn inductor with the experimentally measured coil resistance, with the given geometry and material properties of copper.

In the COMSOL implementation, the physics domains are advanced with a segregated solver that solves the modules in the order of laminar flow, heat transfer, magnetic field and electrical circuit. The outputs of each domain provide the boundary inputs to the next domain at each time step, as parts of a general time dependant solver. Thus, the laminar flow solutions supply the convective velocity field and pressure distribution that enter heat transfer equations. The acquired results from heat transfer then updates the temperature dependent material properties like magnetization and local $B(H,T)$ of the thermomagnetic material, and give the $B$-field distribution. In our transient model, the boundary conditions at inlets switch periodically between hot and cold.

**Figure 3** shows the schematic of the 3D system. In **Figure 3.a**, the device geometry is illustrated, with the hot water inlets from the top, and cold water inlets from the bottom for both sides of the TMG. As starting conditions for our simulations, hot water flows on the left side with no cold water, while the right side receives only cold water. This pattern alternates between both sides at a given frequency. In reality, switching between hot and cold is not perfectly sharp because the switching valves require at least 20 ms to switch the flow. We consider this in our simulations by smoothing the water flow by twice this value. This avoids non-realistic, steep changes in the boundary condition, which also cause numerical solver errors. Our model uses the same control parameters as for the experiment, as summarized in section 3. Furthermore, all performance parameters can be obtained with the same equations (1) to (4), as for the experiment. As only known material properties are used, and the complete geometry is defined, there are no free parameters for the simulations. **Figure 3.b** exemplarily shows a simulated temperature distribution, when the left TMMs are hot and paramagnetic, and the right TMMs are cold and ferromagnetic in the last simulated cycle. In this figure, we can clearly observe that the temperature in the top parts is significantly higher than in the bottom parts of the device, which results from the position of the cold and hot water inlets. This is an indication of undesirable heat transfer to the passive parts and an undesirable temperature gradient over the entire device. In addition, the temperature change after entering the mixing chamber and the non-uniform temperature distribution inside the chamber are visible.

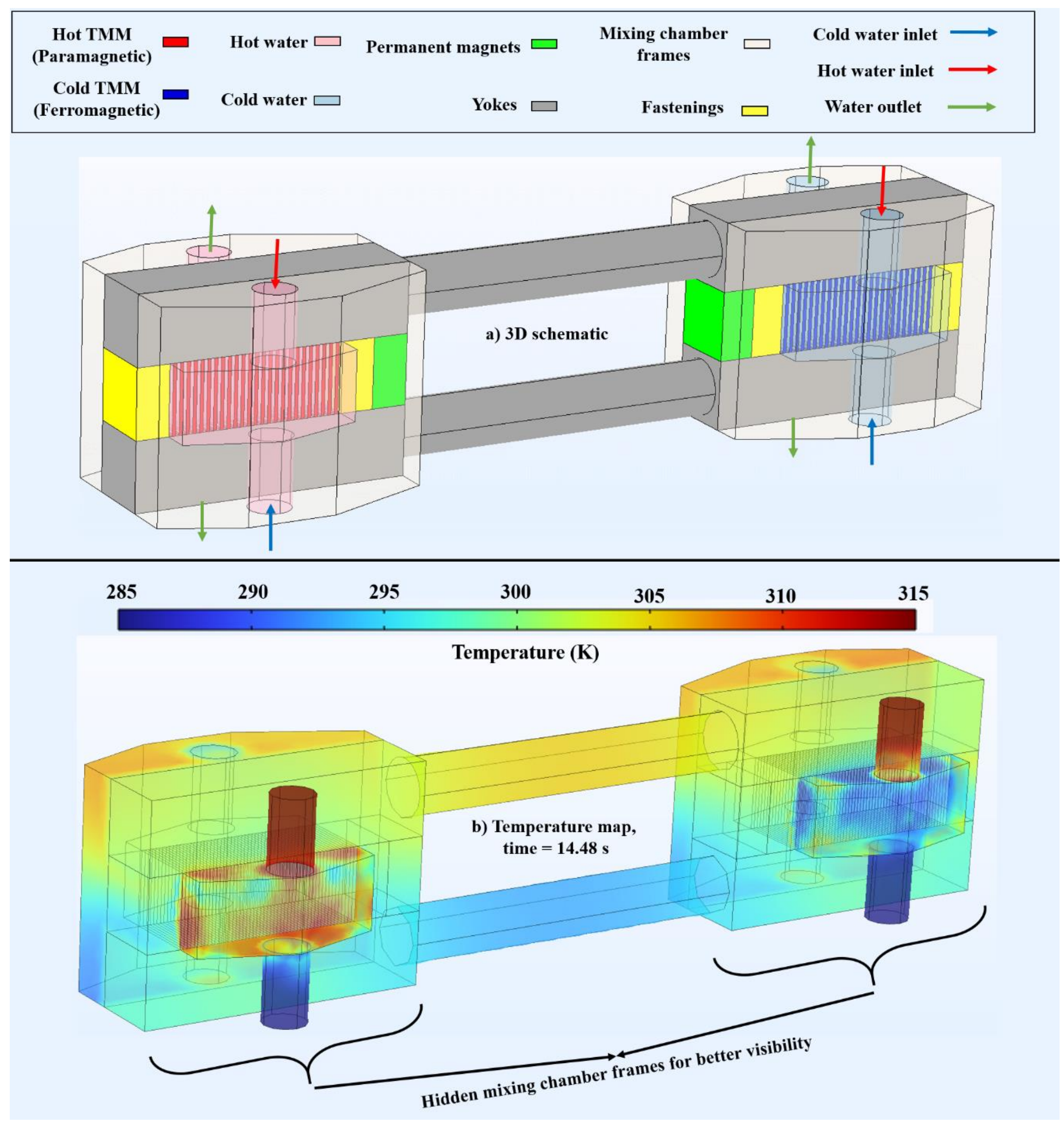

***Figure 3: 3D geometry for Multiphysics simulations and temperature distribution: a)*** *This 3D schematics of the TMG device geometry includes yokes (gray), permanent magnets (green), fastenings (yellow), mixing chamber frames (white). To illustrate the alternating temperature between both sides, the case is shown where hot water (light red) is heating the left plates of TMMs (dark red), and cold water (light blue) cools the right TMMs (dark blue). Hot water for both sides enter from top front (red arrow), cold water from bottom front (blue arrow) and the outlets are on back top and bottom (green arrow)* ***b)*** *Example of simulated temperature distribution while the right part is exposed to cold water and left part is exposed to hot water. Simulation time is t = 14.48 s, within the last simulated cycle, 0.73s after switching the flow. For better visibility of water and TMM plates, temperature gradient in the frames of the inlet mixing chamber is removed.*

To illustrate the advantage of these simulations, we show in **Figure 4** the temperature development of the key components of our TMG during the full 15 seconds of our simulation. The data in this plot represent the average component temperature on each side, including the active TMM plates, top and bottom yokes, magnet, inlet and outlet mixing chamber frames, and also left and right fastenings. Although the TMMs experience the most pronounced temperature change, temperature alternates only between 295 K and 304.4 K, which is way less compared to the change between hot and cold water, alternating between 285 K and 315 K. Already this reduction of $\Delta T$ by a factor of three illustrates that our transient simulations can help to understand the inefficiencies of our TMG. For this, we the consider heat transfer to other components. Whereas heat transfer to TMM plates is intended, heat exchange with other components is unfavourable, as it represents unwanted heat losses. Indeed, the adjacent components, such as the yokes and fastenings, exhibit less gradual temperature changes due to their high thermal mass and limited contact area with the fluid. The temperature change of the permanent magnet is even lower because they are only indirectly exposed to the temperature changes. In contrast, the temperature of left and right fastenings changes by 1 K during each cycle because its contact area to the fluid is much higher than other components. It's worth to note that the temperature of top and bottom yoke exhibit opposite trends. Since the hot water enters the device from top, but the cold water from bottom, the temperature of top and cold yokes at the initialization of the simulations starts closer to the hot and cold water temperatures, respectively. Due to their large thermal mass, these yokes take around 13 seconds to approach the average temperature. That's why for the later detailed analysis we focus on the last simulation cycle, where almost stable cycle conditions are obtained. Besides the main trend in the yokes' temperature, in each cycle additional small fluctuations are observed for both yokes. The reason for these fluctuations is the lack of proper insulation, contributing to low efficiency of the device.

To summarize this section, our simulations allow to access information, which is very hardly obtainable by experiment. In particular measuring the temperature of the TMM in-operando is challenging, as drilling holes for thermocouples into this very brittle material is hardly possible, and thermocouples can affect the heat flow. With simulations, these properties are obtained directly, and averaged over the complete component. Moreover, these simulations give the heat flow to each component during the heating half-cycle, which we use in the section 7 to identify the origin of inefficiency of this TMG.

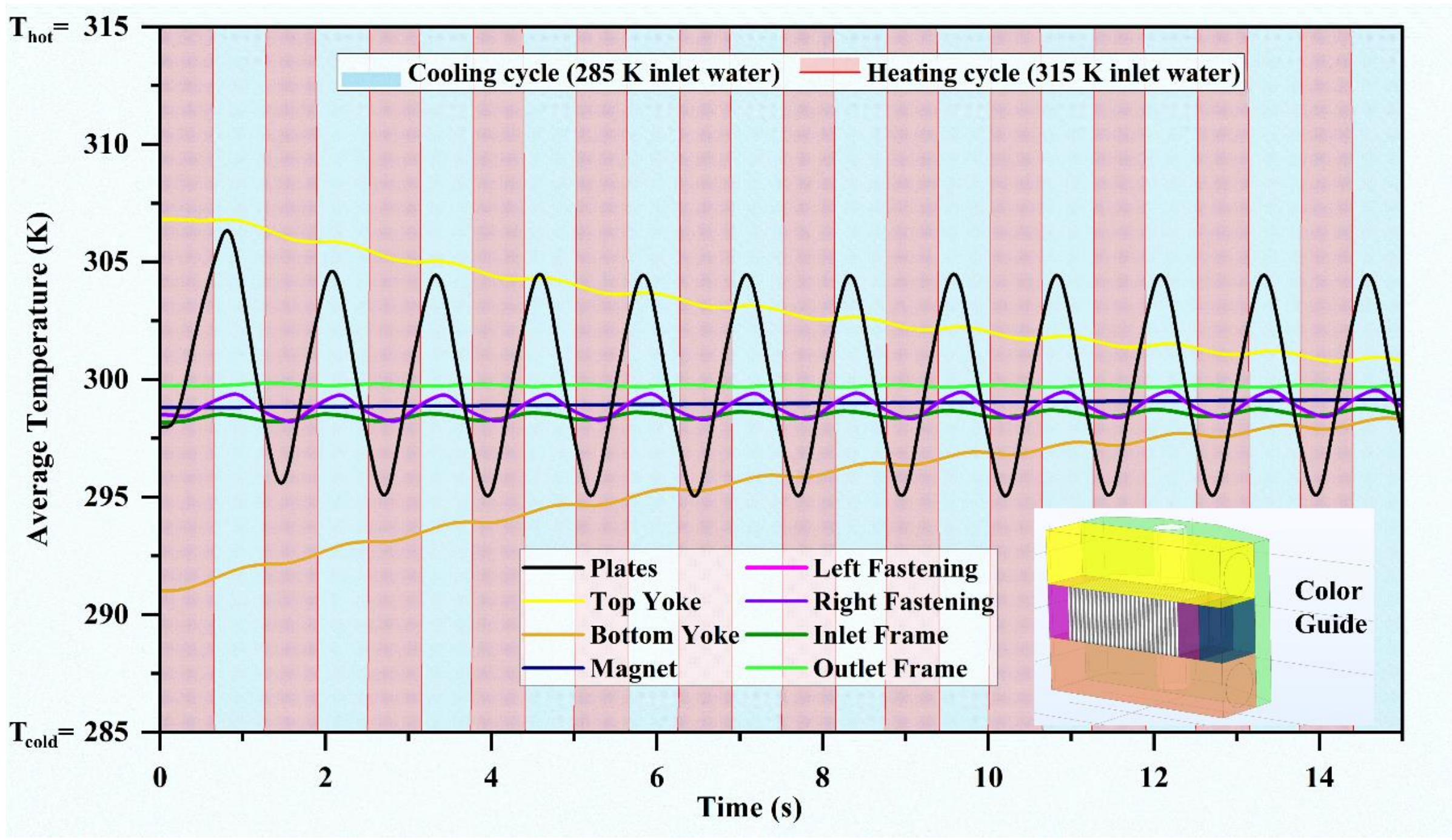

***Figure 4: Time dependent temperature change of key TMG components:*** *Averaged temperature of each component on one side during the first 12 simulated cycles, where hot and cold half cycles alternate, as marked by the red and blue background color. Line colors are selected accordingly to the different device components, as sketched within the insert. During the last cycles also the temperatures of the components at distance to the fluid has stabilized, which allows to use these cycles for detailed analysis. Temperature axis is scaled to the cold and hot fluid temperatures to illustrate that the TMM plates experience a much lower temperature variation.*

# 6. Validation of simulations

Before moving to inefficiencies identification, we validate our simulations with the experimental data of the fully characterized TMG prototype [13]. We focus on the key performance parameters 1) induced voltage and 2) output power at experimentally optimized process parameters.

First, we compare the time dependency of open circuit voltage from our 3D simulation with experiments. **Figure 5.a** shows the last two seconds of our simulation, containing the stabilized heating/cooling cycle. We observe a reasonable agreement for waveform shape, and an excellent agreement for amplitude. The peak open-circuit voltage observed experimentally was 0.165 V, whereas the simulation gives 0.158 V, which is a deviation below 4%. The good agreement of simulations and experiments reveals that our multi-physics model correctly captures the essential physics of the TMG.

Second, we compare the output power under load, which is the most important performance parameter of a TMG. As with the experiments for optimum output power, we connected a resistive load of $R_l$ = 9.2 Ω to our coil, exhibiting a slightly lower DC resistance of $R_c$= 7.7 Ω. **Figure 5.b** shows the instantaneous electrical power obtained by our COMSOL simulations, again for the stabilized, last heating/cooling cycle. The measured experimental peak power is shown by a solid blue line at 0.84 mW, whereas the simulation predicts 0.8 mW, both having a difference of just 5%. The good agreement for this second, independent parameter, is of particular importance, since it addresses power, which is a conservative property like energy. As experiments and simulations agree with both, thermal input power and electric output power, we are therefore confident that also the values for the intermediate energy conversion stages have a reasonable accuracy. As described in the following section, these are only accessible by simulation and give a detailed understanding of complete power flow.

To sum up, the good agreement between experiments and simulations within 5% allows to call our simulations a digital twin - and use it for a detailed analysis of this TMG.

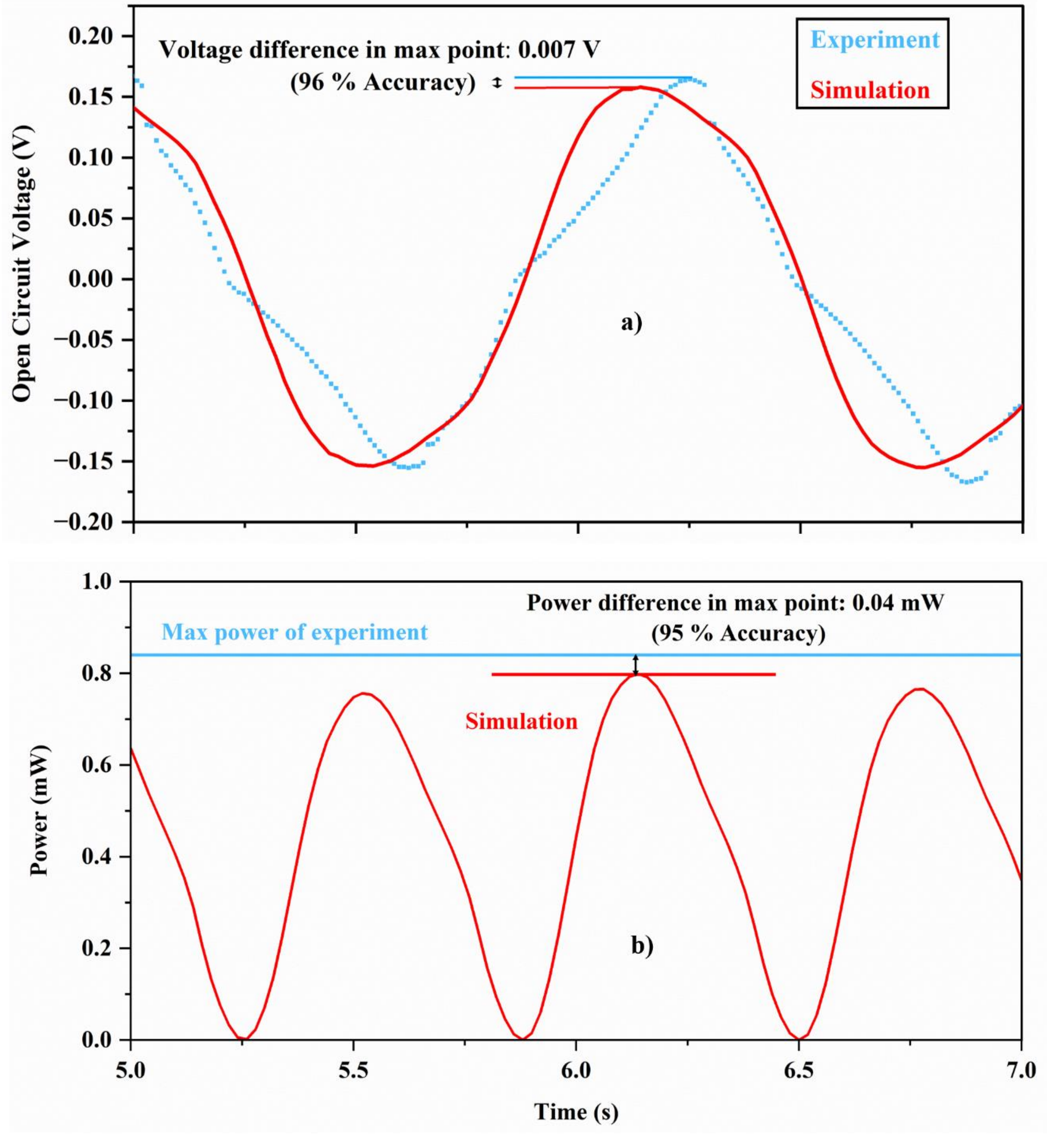

***Figure 5: Validating simulations with experimental results*** *[13]* ***for a) Open circuit voltage and b) Generated power.*** *The measured experimental data is shown by light blue lines while the multi-physics 3D simulation by solid red lines. Simulations agree within 96 % with respect*

*to open circuit voltage and 95 % with respect to output power, which allows to call our simulations a digital twin of the TMG.*

# 7. Identifying origins of inefficiencies

To understand why the system efficiency is much lower compared to the material efficiency, we summarize the different heat flows within our TMG in a Sankey diagram (**Figure 6**). This diagram quantifies the distribution of input thermal power to the different losses, and useful electrical output power, offering insight into the device's energy conversion performance. This diagram follows the hot fluid pathway and reveals how much energy is wasted in each part, including the mixture of hot and cold water from previous cycles.

To calculate this Sankey diagram, we use the heat flows for each component during the last, stabilized heating/cooling cycle of the simulation (from 13.75 s to 15.00 s). Our calculations consider both, the cyclic nature of this process, and that the components of the device do not absorb heat simultaneously, as analysed in section 8. As described in more detail in Supplementary Note 4 and Figure S5, we track for each component the specific time interval during which this component absorbs heat, and the integral over one cycle gives the corresponding heat flow. To obtain an idea on the accuracy of our approach, we also analysed the release of heat, which should be same. The error given within the Sankey diagram for the heat flows represents the difference of absorption and release. Except for the minor losses of parts far away from the fluid, these errors are acceptable low.

In the following we will shortly describe all heat flows, discuss their origin and possible routes to improve system efficiency. We start with the simplest energy line of the Sankey diagram and end with the most difficult one:

a) The total input thermal power by hot water during the heating half-cycle is 801 W, as directly obtained by Equation (3).
b) The alternating hot and cold water first reaches a mixing chamber, which distributes the water to the TMM plates. This design is detrimental for efficiency, since mixing hot and cold water just gives water at a medium temperature, which simply annihilates energy without use. Indeed, at this initial step already 252 W is lost, which makes this mixture one of important sources of inefficiency in this device.
c) In the mixing chamber an additional loss of 10 W occurs by the heat exchange to the air through the frames of mixing chamber. Though the mixing chambers are made from plastic with a low thermal conductivity, their large contact area gives makes this contribution significant.
d) Thus, a power of 539 W reaches the plates channels totally. This illustrates that removing this mixing chamber in future designs will allow to increase efficiency.
e) Heat dissipation in the yokes and fastenings contribute with losses of 14 W and 2W, respectively. The heat transfer into the yokes is more than fastenings due to higher thermal conductivity of steel and also their larger heat transfer area.
f) A large portion of the fluid heat, around 435 W, is lost as unused outlet water energy. Origin of this loss is discussed below.
g) Finally, 87 W go into the TMM plates, which is around 11% of the input energy. From this amount only $4.6 \times 10^{-4}\ W$ is converted into the electric power. This deserves a detailed discussion, given below.

This energy flow analysis highlights key heat loss mechanisms and parts, and also provides a quantitative base for identifying opportunities to enhance device performance. First, we discuss the power, which is lost, before it even reaches the TMM. At this first stage, several approaches to improve device efficiency are quite evident: preventing mixture of water (b), improving thermal insulation (e), and reducing parasitic conduction to passive components (c,e). A bit more complex is f), which occur due to the high cycle frequency, which makes hot water leaving the TMG unused. We attribute this to the simplified cycle run by this TMG. In the ideal case, a TMG should follow a 4-stage Ericcson cycle, as described e. g. by Dzekan et al [14]. But in this TMG the temperature was only alternated between hot and cold. Thus, it is more a two-stage cycle, without completing the isofield heat exchanges. For future improvement, we thus propose to interrupt the fluid flow before switching between hot and cold flow, which should allow to complete the heat exchange between fluid and TMM. However, it's worth to note, that by principle not both is possible: Reaching a high efficiency, requiring a lot of time for complete heat exchange; and reaching a high output power, requiring a short cycle time, leaving heat exchange incomplete. And here we analyse process conditions optimized for high output power[13]. To sum up this section, only 11 % of the thermal input power reaches the TMM plates. This gives quite some room for future improvements, but for maximum output power, some losses are unavoidable.

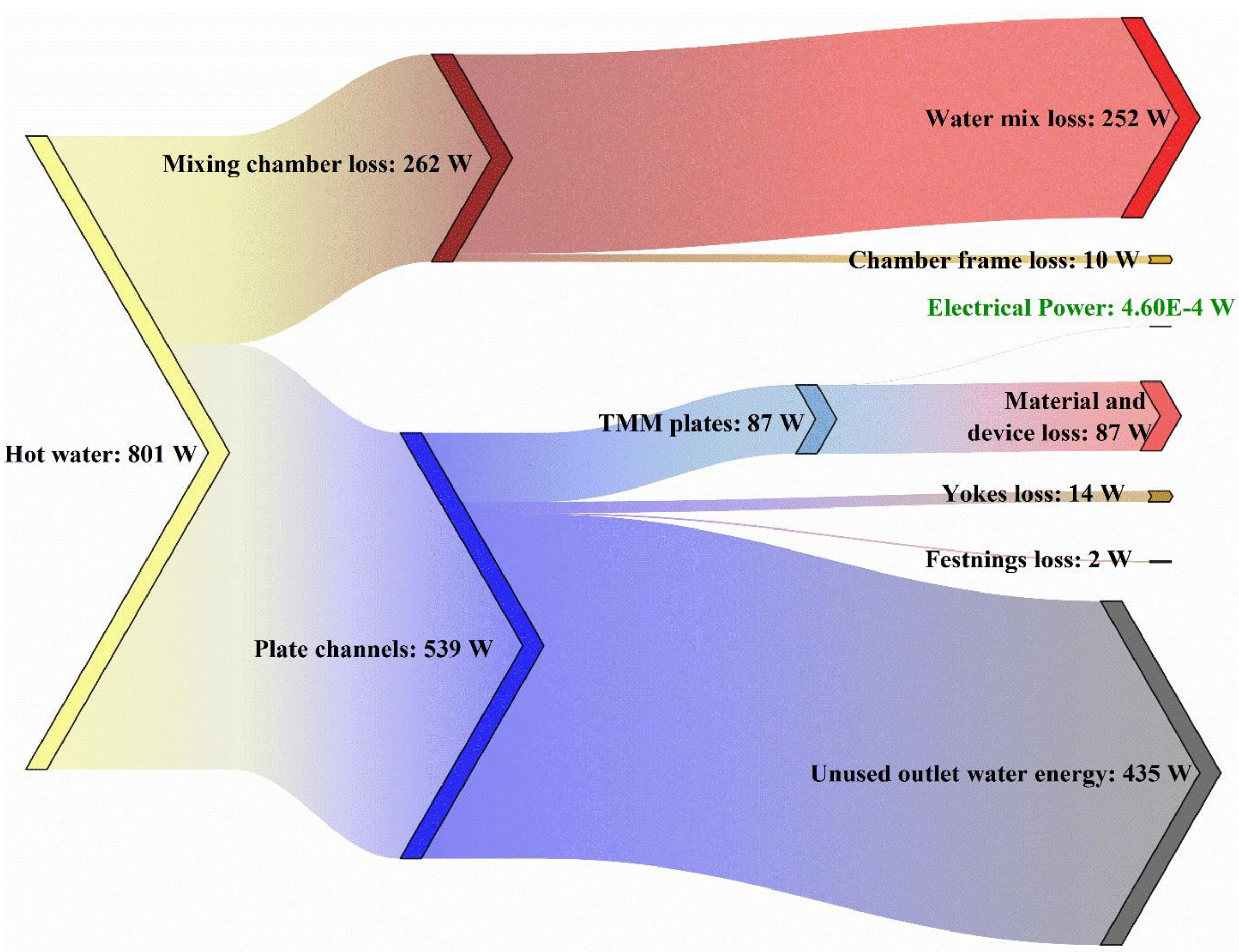

***Figure 6: Energy flow through the device as Sankey diagram.*** *This plot illustrates how the input power by hot water during a heating cycle is distributed within the TMG. It identifies losses through different, non-active parts of the device, which are in direct contact with water such as mixing chamber frames, yokes, and fastenings, and waste energy due to mixing, until*

*finally a small amount of useful power is delivered to the TMM plates and an even smaller fraction is converted to electric power.*

As next step, we discuss the conversion of magnetic to electric energy, which according h) reduces the power of 87 W entering the TMM plates to $4.6 \times 10^{-4}\ W$, leaving the TMG as electrical power. To put these numbers in a context, we conder the materials efficiency, described within the introduction. However, we consider the previous section, which effectively reduces the temperature amplitude of the TMM from 30 K between hot and cold inlet to about 10 K (**Figure 4**). Thus, we must consider Carnot efficiency as the thermodynamic limit for the temperature span of just 10 K, which is just 3%. When considering the materials efficiency of 55% of Carnot, as described within the introduction, this reduces the maximum efficiency of a TMG to 1.5%. When taking the power flow to the TMM plates of 87 W, still an electric output of about 1.3 W should be feasible, but only $4.6 \times 10^{-4}$ W is obtained by both, experiments and simulations. This large gap is the key to future improvements of TMGs, which therefore deserves a bit more discussions. First, we consider that the input power to the TMM is an integral value, but local magnetisation decides on switching the magnetic flux. In other words, some cold paramagnetic regions can strongly inhibit the magnetic flux, even when most of the TMM already switched. Thus, it is decisive to avoid inhomogeneities, which are analysed in Section 8. Second, magnetisation is a volume property, but a magnetic flux goes through an area. This means, that the energy, required to change the magnetisation within the volume is not directly related to the total flux change, which decides on the output power. In other words, there is plenty of room to improve the conversion step from magnetic to electric energy. This includes redesigning the electric circuit and coil configuration, designing an efficient magnetic circuit with minimal stray fields and short circuits, and improving material properties, such as lower heat capacity of the thermomagnetic plates and a sharper magnetization change of the active material with temperature. This example illustrates that the development of TMG, which up to now was mostly driven by physicists and material engineers, should include now also mechanical and electrical engineers.

# 8. Origin of low cycle frequency

In cyclic power generators, the average electrical power $P$ is the energy $E$ converted per cycle multiplied by the frequency $f$:

$$P = E \times f \quad (5)$$

Therefore, in an ideal TMG all TMM should heat and cool immediately and completely after switching between cold and hot fluid. Identifying delays compared to this ideal situation, allow to understand the factors limiting cycle frequency. For this, we will follow the flow of the heat through the system in time and examine when the temperature peaks. In addition, we will examine local inhomogeneities in temperature, as they can reduce conversion efficiency.

For our analysis of heat propagation, we choose the final heating half cycle, as this reached stable conditions. **Figure 7** shows the time dependency of the temperatures of the main components together with their location in one side of the TMG. Control parameter is the water inlet temperature, marked in yellow, which switches from cold to hot at 13.79 s. First, we analyse the average water temperature within the mixing chamber (black line), which starts to increase immediately after switching, and reaches its maximum temperature soon after switching back to the cold fluid at 311.4 s. At no time a stable temperature is reached within the mixing chamber. The associated losses within the mixing chamber become evident by it reduce maximum temperature of 311 K compared to the inlet temperature of 315 K - the average temperature is even lower. Second, we analyse the part of the TMM plates, which are located closest to the fluid inlet, which we name “starting face of TMM”, and mark in green. These follow the temperature within the mixing chamber and reach almost the same maximum temperature, after a delay of just 0.04 s. Third, we analyse the temperature at the end of the plate with respect to the fluid flow, called “ending face of the plates”, marked in blue. We observe an additional delay of 0.18 s compared to the front face, which mostly originates from the time the water requires to flow through the channels. During this flow, the water exchanges heat with the plates, and accordingly the maximum temperature of the ending face is 5 K lower compared to the front face. Thus, there is a substantial inhomogeneity with the plates in respect to both, temperature and switching time. Indeed, when comparing the starting and ending faces of the plates, the ending face still heats up, while the starting face already cools down. This is detrimental for efficiency since only a fraction of the plate switches the magnetic flux. Moreover, a cold fraction of the plates can act as a ferromagnetic “shortcut” for the magnetic flux, bypassing the induction coils. Forth, the average temperature in the outlet mixing chamber is shown in red. It reaches its maximum of 304.8 K, about 0.1 s after the plates' outlet faces. This marks the completion of the heat propagation through the TMG.

This path of heat flow identifies the key limitations for frequency. The mixing chamber is the most limiting factor, as the water requires time to travel through it, which delays the time before the TMM plates heat up. This is a second reason in addition to mixing losses, why future design of TMG should avoid a mixing chamber. In addition, the fluid flow through the channels is limiting frequence. As these channels also result in inhomogeneous temperatures within the plates, it would be beneficial to avoids fluid flow through a TMG. Indeed, for thermoelastic harvesters recently a novel fluid management system was proposed, which could minimize these effects significantly [30], but a design transfer from the SMA wire used there to TMM plates used here, is missing.

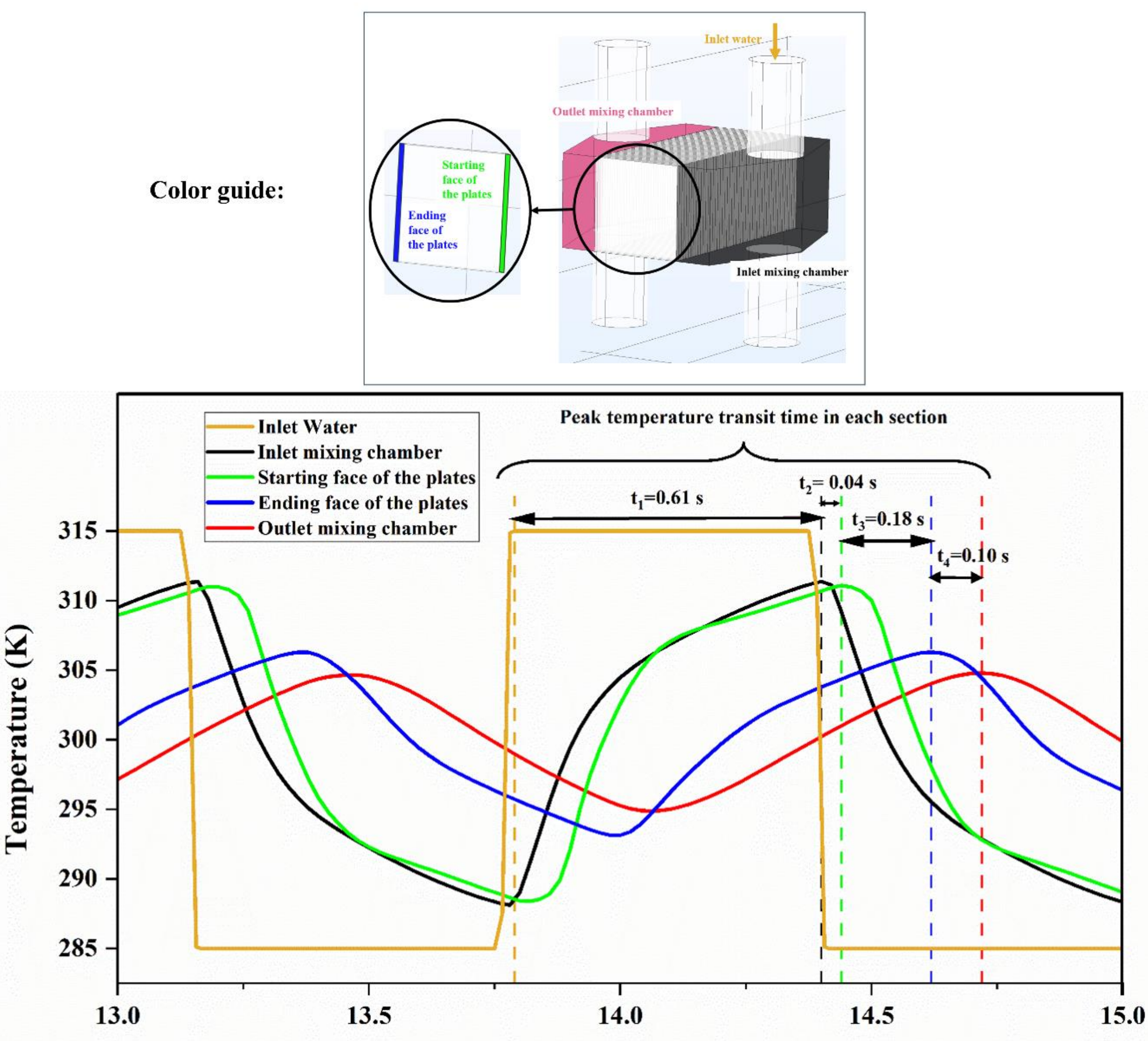

***Figure 7: Heat propagation during a heating cycle to identify frequency limitations of the TMG.*** *Time dependency of temperature of the key components within the TMG, as marked by different colors and sketched in its color guide. After switching from cold to hot water at t = 13.79 s, each component requires a specific time until it reaches its maximum temperature. This time is marked in this graph by dashed horizontal lines, and the delay with respect to the previous component gives an idea, how much this component limits cycle frequency.*

This heat flow analysis revealed substantial temperature inhomogeneities within the TMG, which are detrimental for its performance. These are not visible within the previous section, which averages the heat flow to each component in time and space. Thus, in the following we will expand our analyse to local temperature inhomogeneities, including their time dependency, and discuss their detrimental effect on efficiency. The associated inhomogeneous distributions of the magnetic flux are shown in supplementary Figure. S6. As reference for the following analysis, we consider an ideal TMG, where all TMM changes on each side changes its temperature homogenously, switching the magnetic flux completely between hot and cold side, as sketched in **Figure 2**.

The temperature distribution across the TMM plates in full 3D are shown in **Figure 8** for the heating cycle. For clarity only three representative TMM plates depicted: the central plate, and those located at the left and right ends. The temperature evolution at $t$ = 13.96 s, when the average plate temperature starts to rise, is shown in **Figure 8.a**. As the inlet is located closest to the centre plate, the hot water first reaches the inlet face of the central plate. Accordingly, temperature increase only at this location, while the other parts of the middle plate and also the complete adjacent plates remain at a lower temperature. The following intermediate times at $t$ = 14.2 s and $t$ = 14.4 s are shown in **Figure 8.b** and c, respectively. At both times the inhomogeneous temperature along the water flow direction parallel to the $y$ direction becomes more pronounced. The coexistence of both, hot and cold regions reduce the total flux change in this side. Accordingly output power is reduced compared to the ideal TMG, sketched at the end of last paragraph. Moreover, these figures also reveal that the temperature along the $z$-direction is inhomogeneous. The bottom part of the plates is much hotter than the top part, which originates to the location of the hot inlet, placed at the top. As the hot TMM regions are paramagnetic, they block the magnetic flux through the cold regions. This inhomogeneity clearly illustrates the need to consider the local temperature and not only the average temperature. Finally, in **Figure 8.c** the temperature at $t$ = 14.58 s is depicted, which is the time, when the maximum temperature, averaged over all plates, is reached. Though the average temperature increased, most of the temperature inhomogeneities persist, revealing that these origins of inefficiencies are relevant throughout a complete cycle. Moreover, the front face of the middle plate already cools down and becomes ferromagnetic. Thus, this part already conducts part of the magnetic flux, despite that the cold side should block the magnetic flux. The detrimental effect of this inhomogeneity becomes evident, when considering the principle of this TMG, as sketched in **Figure 2**. This cold region acts as a shortcut for the magnetic flux, bypassing the induction coils and therefore not contributing to output power.

To sum up, this time resolved local analysis identifies temperature inhomogeneities as origin why from the 87 W reaching the TMM plates only $4.6\times10^{-4}$ W is converted to electricity, as summarized in the Sankey diagram (**Figure 6**). Coexistence of hot and cold regions along the water flow direction reduce the total amount of flux switched. Coexistence of cold regions on the hot side act as a kind of shortcut, allowing the flux to bypass the induction coils, not contributing to output power. Coexistence of hot and cold regions along the magnetic flux direction reduce the total flux, as the flux is blocked by the hot regions.

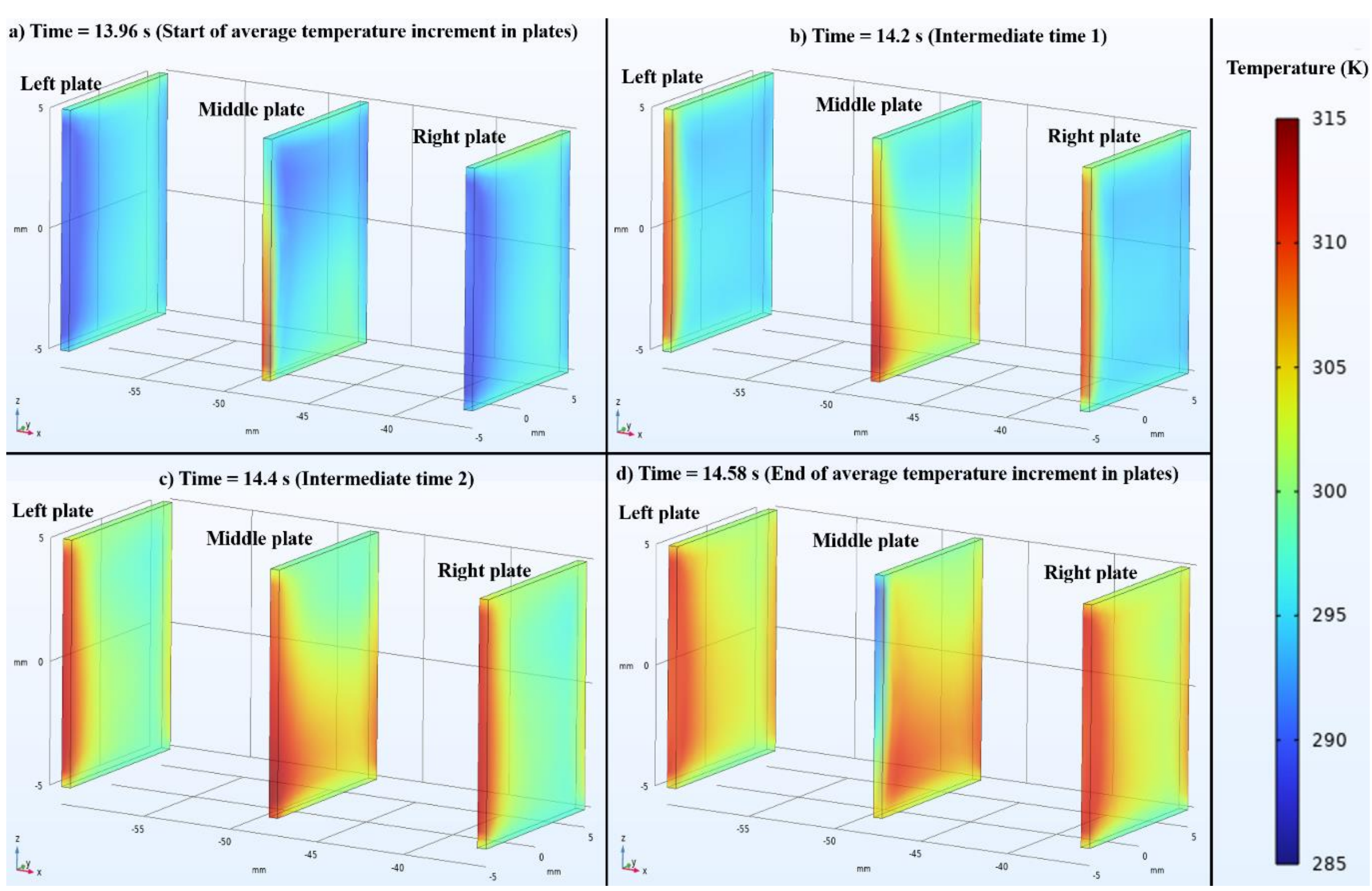

***Figure 8: Identifying inhomogeneous temperatures within TMM plates as origin of inefficiencies.*** *This figure exemplarily depicts three TMM plates, one in middle and two in left and right ends in four different delay time after switching from cold to hot water at t = 13.75 s. Inhomogeneities occur along the fluid flow direction, between middle and outside plates, and between top and bottom of the plates, since the hot and cold inlets are on top and bottom, respectively.*

# 9. Conclusion

Thermomagnetic generator (TMG) is an emerging technology for harvesting low-grade waste heat, but today's systems suffer from low system efficiency and low cycle frequencies. To understand the origins of these shortcomings, we develop a digital twin of the TMG reaching the highest power density today [13]. In brief, the key progress in understanding TMGs can be summarized as follows:

- **Digital Twin:** Only fully coupled 3D simulation allow to describe a TMG appropriately, since decisive processes like magnetic flux and fluid flow are perpendicular to each other. 2D models do neither describe magnetic stray fields nor heat flow appropriately. Our 3D Multiphysics model only requires known material parameters, geometry and process parameters as input parameters, and give the complete time dependency of temperature, magnetic flux for each position, as well as output voltage and power.
- **Experimental Validation:** Our digital twin gives the experimentally measured [13] open circuit voltage and power output with an accuracy of 96% and 95%, respectively. This high agreement for both most important performance parameters of a TMG allows to use this digital twin for an in-depth study.
- **Energy Flow and Losses:** Our digital twin gives a Sankey diagram, containing all energy flows and loss mechanisms. Only 11 % of the input thermal energy reaches the TMM plates, and a much lower fraction is converted to electric energy. Our analysis identifies the reasons for the large difference between system and material efficiency, and paves the way for increase system efficiency by design optimization.
- **Cycle Frequency:** By following the heat flow through the TMG in time, our digital twin identifies the origins of low cycle frequency, which must be overcome for high output power. Moreover, the detailed information on temperature distribution within the TMM allows to identify temperature inhomogeneities as origin of low system efficiency.

Though this work focused on understanding the performance of a TMG, it also gives clear guidelines for their improvements. First, no mixing chamber should be used, as mixing reduces both, efficiency and frequency. Second, any temperature inhomogeneities within the TMM should be avoided, as they reduce the magnetic flux change strongly. As our digital twin is validated experimentally, it can be also used for the development of the next generation of TMG. We expect that this will accelerate the development of TMGs with higher efficiency and frequency strongly.

# ACKNOWLEDGMENTS

The authors would like to express their sincere gratitude to Dietmar Berger for his valuable assistance and contributions to this work. The authors gratefully acknowledge that this project has received funding from the European Union under grant agreement No 101119852. We gratefully acknowledge compute time and data storage on the vlsmath1 system provided by the Department of Information Services and Computing at HZDR.

# AUTHOR CONTRIBUTIONS

**Ali Izadi**: Conceptualization (equal); Data curation (lead); Formal analysis (lead); Investigation (lead); Methodology (equal); Software (lead); Validation (equal); Visualization (equal); Writing – first draft (lead); Writing – review & editing (equal).
**Bruno Neumann**: Visualization (equal); Writing – review & editing (equal); Conceptualization (equal).
**Sebastian Fähler:** Conceptualization (equal); Funding acquisition (lead); Methodology (equal); Project administration (lead); Resources (lead); Supervision (lead); Writing – review & editing (equal).

# DATA AVAIBILITY STATEMENT

Data and metadata that support the findings of this study are openly available in RODARE at DOI:10.14278/rodare.4553. This data includes material properties used for COMSOL simulation and result data shown in plots. As the COMSOL uses a priority data format, which does not follow FAIR principles, these files are not included.